\documentclass[letter]{aa} 
\usepackage[authoryear]{natbib}

\usepackage{graphicx}
\usepackage{amsmath}
\usepackage{amssymb}

\usepackage{txfonts}
\begin{document}
\title{\textsl{INTEGRAL} and \textsl{Swift} observations of EXO 2030+375 during a giant outburst\thanks{ Based on observations with INTEGRAL, an ESA project with instruments and science data centre funded by ESA member states (especially the PI countries: Denmark, France, Germany, Italy, Switzerland, Spain), Czech Republic and Poland, and with the participation of Russia and the USA.}}


\author{D. Klochkov\inst{1}, D. Horns\inst{1}, A. Santangelo\inst{1}, R. Staubert\inst{1}, A. Segreto\inst{3}, C. Ferrigno\inst{3}, P. Kretschmar\inst{4}, I. Kreykenbohm\inst{1,5}, A. La Barbera\inst{3}, N. Masetti\inst{8}, M. McCollough\inst{2}, K. Pottschmidt\inst{6}, G. Sch\"onherr\inst{1}, J. Wilms\inst{7}}

\institute{
  Institut f\"ur Astronomie und Astrophysik, University of T\"ubingen, Sand 1, 72076 T\"ubingen, Germany 
\and
 Smithsonian Astrophysical Observatory, 60 Garden Street, Cambridge, MA 02138, U.S.A
\and
  INAF IFC-Pa, via U. La Malfa 153, 90146 Palermo, Italy
\and
 Integral SOC ESA, Madrid, Spain
\and
 INTEGRAL Science Data Centre, Chemin d'Ecogia, 16, 1290, Versoix, Switzerland
\and
 Center for Astrophysics and Space Sciences, University of California, San Diego, La Jolla, CA 92093-0424, USA
\and
 Dr. Karl Remeis-Sternwarte, Astronomisches Institut, Universit\"at
Erlangen-N\"urnberg, Sternwartstr. 7, 96049 Bamberg, Germany
\and
  INAF IASF-Bo, via Gobetti 101, 40129 Bologna, Italy
}

   \date{}

  \abstract
  {}
   {We investigate the X-ray spectral and timing properties of
the high mass X-ray binary \object{EXO 2030+375} observed during its
June--September 2006 giant (type II) outburst. 
}
   {The data analyzed in this work are
from partly simultaneous observations with \textsl{INTEGRAL}
and \textsl{Swift}.
The pulse period $P$ and its temporal derivative $\dot P$ are
measured. X-ray pulse profiles in different energy ranges and time
intervals are constructed. Pulse averaged X-ray spectra for different
time intervals are studied.
}
   {We report a strong spin-up of the source during the
outburst, comparable to that observed in 1985 during the previous
giant outburst when the source was discovered.
The value of $\dot P$ is found to be linearly related to the X-ray luminosity
of the source during the outburst. 
For the first time the hard X-ray ($>25$~keV) characteristics of 
the source during a type II outburst are studied.
The X-ray pulse profiles apparently change with luminosity. 
The X-ray spectral continuum in the 3--120 keV energy range is modeled with an
absorbed power law with an exponential cutoff around $E\sim26$~keV.
An iron emission line at $\sim$6--7 keV is observed.
The spectrum reveals some features between 10 and 20 keV which
can be modeled either by a broad emission line at $\sim$13--15 keV
(a ``bump'') or by two absorption lines at $\sim$10 and $\sim$20 keV.
} {}

   \keywords{X-ray binaries; neutron stars; accretion disks}

\authorrunning{Klochkov et al.}
\titlerunning{EXO 2030+375 in giant outburst}
   \maketitle
%

\section{Introduction}

Discovered by \textsl{EXOSAT} in 1985 during a giant outburst
\citep{Parmar_etal89a}, the transient accreting pulsar EXO 2030+375
belongs to a high mass X-ray binary system with a B0 Ve star as
optical companion \citep{Coe_etal87}. The orbital period and
eccentricity are $\sim$46\,d and $\sim$ 0.42, respectively
\citep{Wilson_etal05}. According to the generally accepted model for
Be/X-ray binaries, the X-ray emission in such systems
during outbursts is produced
when the compact object accretes from a quasi-Keplerian disk around
the equator of the rapidly rotating Be star. Such a mechanism explains
normal (type I) outbursts with X-ray luminosity
$L_{\rm X}\lesssim 10^{37}\,\text{erg}\,\text{s}^{-1}$,
associated with the periastron passage of
the compact object. Sometimes, however, Be/X-ray binaries show giant
(or type II) outbursts with X-ray luminosity $L_{\rm X}\gtrsim
10^{37}\,\text{erg}\,\text{s}^{-1}$. Such outbursts are thought to be due to a
dramatic expansion of the disk surrounding the Be star, leading to
the formation of an accretion disk around the compact object
\citep{Coe00}.

During the 1985 giant outburst the X-ray luminosity of the source
reached a value of
$L_{\rm 1-20~keV}\sim 2 \cdot 10^{38}\,\text{erg}\,\text{s}^{-1}$
(for a distance of 7.1 kpc, \citealt{Wilson_etal02}). The
spin period of the pulsar ($\sim 42$\,s) changed dramatically, with
a spin-up time scale $-P/\dot P \approx 30$\,yr
\citep{Parmar_etal89a}.

\begin{table}
\caption{Summary of observations.}
\label{obslist}
\centering
\begin{tabular}{c c c c c}
\hline\hline
Obs. & Instrument & Obs.time  & Mean & Mean ASM        \\
     &            &  (ks)     &  MJD & flux (mCrab)    \\
\hline
OBS1        & ISGRI+JEM-X  & 62  & 53942.9 & 500 \\
OBS2        & ISGRI+JEM-X  & 140 & 53967.6 & 500 \\
OBS2        &  XRT + BAT    & 6.3 & 53967.3 & 500 \\
OBS3        &  XRT + BAT    & 6.4 & 54002.5 & 160 \\
\hline
\end{tabular}
\end{table}

From the nearly continuous monitoring of the source with the Burst
and Transient Source Experiment (\textsl{BATSE}) on the Compton
Gamma Ray Observatory (\textsl{CGRO}) and the All Sky Monitor (\textsl{ASM})
on the Rossi X-ray Timing Explorer (\textsl{RXTE}) it is known
that the pulse period remained roughly constant for about a year after the
beginning of the monitoring in 1991, followed by 2 years of
relatively slow spin-up and 6 years of slow spin-down
\citep{Wilson_etal02, Wilson_etal05}. After 2002 a transition to a
global spin-up along with an overall brightening of the normal
outbursts was reported by \citet{Wilson_etal05}.

The giant outburst of EXO 2030+375 presented here 
\citep{CorbetLevine06,Krimm_etal06,McCollough_etal06}
is the second of its kind since the discovery of the source.
The X-ray luminosity at the maximum of the outburst reached
$L_{\rm 1-20~keV}\sim 1.2 \cdot 10^{38}\,\text{erg}\,\text{s}^{-1}$ 
\citep{WilsonFinger06}. In
this \emph{Letter}, based on \textsl{INTEGRAL} and \textsl{Swift} data,
we investigate in $\S$~2.1 the pulse period and the spin-up rate of
the source and in $\S$~2.2 we construct X-ray pulse profiles for
different energy ranges and time intervals. To characterize the
broad-band spectral behaviour we perform in $\S$~2.3 the
spectral analysis
of pulse-phase averaged spectra of the source. A detailed study
of the timing and spectral behaviour as a function of the luminosity
is beyond the scope of this \emph{Letter} and will be presented in a
forthcoming paper.


\section{Observations and data analysis}

EXO 2030+375 was observed by \textsl{INTEGRAL}
\citep{Winkler_etal03} on 27 July and 19--21 August 2006 and by
\textsl{Swift} \citep {Gehrels_etal04} on 19--20 August and 23--25
September 2006. Another simultaneous observation by
\textsl{INTEGRAL} and \textsl{Swift} was performed on 6--8 October
2006, but the data were not available at the time of writing this
\emph{Letter} and will be presented later.  For our analysis we used
the data obtained with the instruments \textsl{IBIS/ISGRI}
(20--300\,keV, \citealt{Ubertini_etal03}) and \textsl{JEM-X} 
(3--30\,keV, \citealt{Lund_etal03}) onboard 
\textsl{INTEGRAL} as well as \textsl{XRT}
(0.2--10\,keV, \citealt{Burrows_etal05}) and \textsl{BAT}
(15--150\,keV, \citealt{Barthelmy_etal06}) of \textsl{Swift}. For
the analysis of \textsl{ISGRI} and \textsl{JEM-X} data the Off-line
Science Analysis (OSA) software (version 5.1) was used
\citep{Courvoisier_etal03}. 
Analysis of \textsl{XRT} data was done with
the standard \textsl{HEADAS} package (version 6.0.4).

\begin{figure}
\centering
\includegraphics[width=8.5cm]{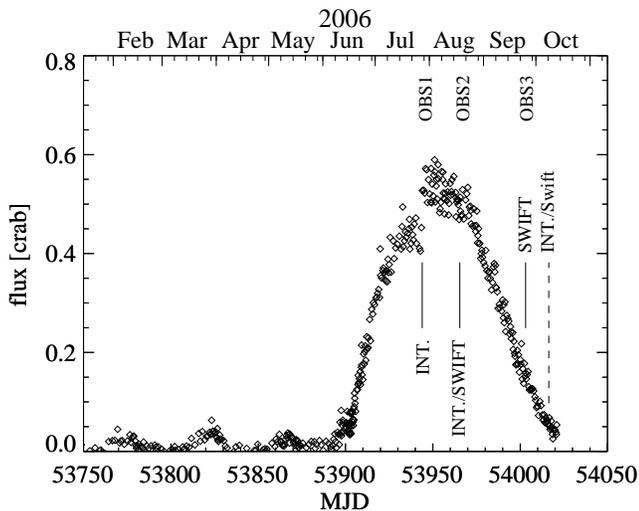}
\caption{The ASM light curve of EXO 2030+375. The times of \textsl{INTEGRAL}
(INT.) and \textsl{Swift} observations analyzed in this work are marked by
solid lines. The dashed line shows another simultaneous observation by
\textsl{INTEGRAL} and \textsl{Swift} for which the data are not yet available.}
\label{asm}
\end{figure}

Table~\ref{obslist} contains
the summary of the observations analyzed in
this work. A part of the \textsl{ASM} light curve\footnote{We used 
the results provided by the ASM/RXTE teams at MIT and 
at the RXTE SOF and GOF at NASA's GSFC.}
of the source including three normal outbursts and the 2006 
giant outburst is presented in Figure~\ref{asm}.
The observations analyzed in this
paper are marked by solid lines and consist of three sets of
pointings. The first set was done by \textsl{INTEGRAL}. The second set
contains simultaneous observations by \textsl{Swift} and
\textsl{INTEGRAL}. The last set of observations was performed by
\textsl{Swift}. These three sets will be referred to as OBS1, OBS2,
and OBS3 throughout the paper. OBS1 and OBS2 observations were made
when the source was at approximately the same luminosity level,
before and after the maximum of the outburst. OBS3 set was made
during the decay of the outburst when the luminosity of the source
dropped by a factor of $\sim$3.

\begin{figure}
\centering
\includegraphics[width=9cm]{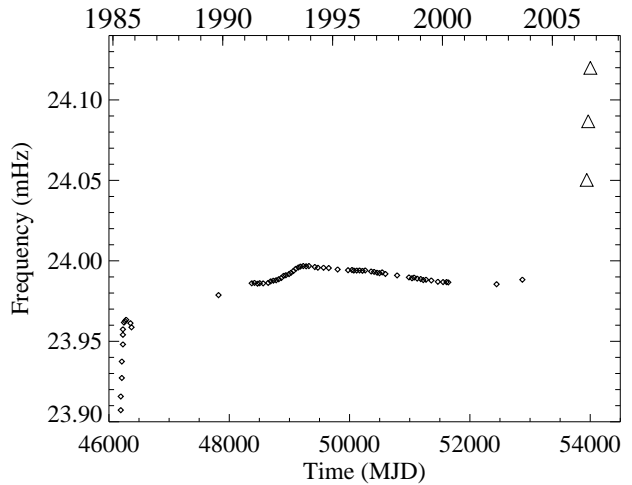}
\caption{Spin frequency of EXO 2030+375. Small diamonds show the spin
frequency development before the 2006 giant outburst (they are taken from 
\citealt{Wilson_etal02,Wilson_etal05}). The large triangles correspond to
our pulse frequency measurements during the outburst. The uncertainties
of measurements and duration of each observation are smaller than the
size of symbols in the plot.}
\label{per}
\end{figure}

\subsection{Pulse period behaviour}

\begin{table}
\setlength{\tabcolsep}{0.15cm}
\caption{Pulse period development. The uncertainties in parentheses (68\%) refer to the last digit(s).}
\label{tabper}
\centering
\begin{tabular}{c c c c c}
\hline\hline
Obs. & MJD  & $P$ & $-dP/dt$       & Flux (3--10 keV) \\
     &      & [s] & [$10^{-8}$ s/s]& [$10^{-9}\,\text{erg}\,\text{cm}^{-2}\,\text{s}^{-1}$] \\
\hline
OBS1 & 53942.7714 & $41.57958(2)$ & $2.93(12)$  & 9.61(6)  \\
OBS2 & 53967.4260 & $41.51706(5)$ & $3.29(19)$  & 10.41(4) \\
OBS3 & 54002.3877 & $41.45954(4)$ & $0.67(10)$  & 3.25(3)  \\
\hline
\end{tabular}
\end{table}

For the timing analysis all times were translated to the 
solar system barycenter and corrected for orbital motion in the
binary. The pulse periods and associated derivatives were determined
for each of the three observation periods individually (Table~\ref{tabper}). 
These values were found by employing initial epoch-folding and a 
subsequent phase connection analysis similar to \citet{Ferrigno_etal06} 
using well defined pulse profiles from a sufficiently
large number of pulses. Variation in pulse shape inside each observation 
is marginal and does not affect our method.

Fig. 2 displays our points together with the historical pulse period 
development as presented by \citet{Wilson_etal02,Wilson_etal05}. The three datasets are consistent with a global solution (valid between
MJD 53943 and MJD 54003) with the following parameters:
$P = 41.47268(6)$ s, $\dot P = -1.627(5)\cdot 10^{-8}$ s/s and
$\ddot P = 4.058(25)\cdot 10^{-15}\,\text{s}^{-1}$ 
for the reference MJD 53992.25 (TDB). The uncertainties in 
parentheses (68\%) refer to the last digit(s). The characteristic
spin-up time scale, $-P/\dot P$, is $\sim$40 years, somewhat longer 
than during the first giant outburst. We also note that the $\dot P$ 
appears to be linearly correlated with the X-ray luminosity
(see last two columns in Table~\ref{tabper}). The formal correlation 
coefficient is 1.0 and the slope from a linear fit is 
$36\pm2\,(\text{s\,s}^{-1})/(\text{erg cm}^{-2} \text{s}^{-1})$, although
it is based on three data points.

\subsection{Pulse profiles}

\begin{figure}
\centering
\includegraphics[angle=-90,width=8.7cm]{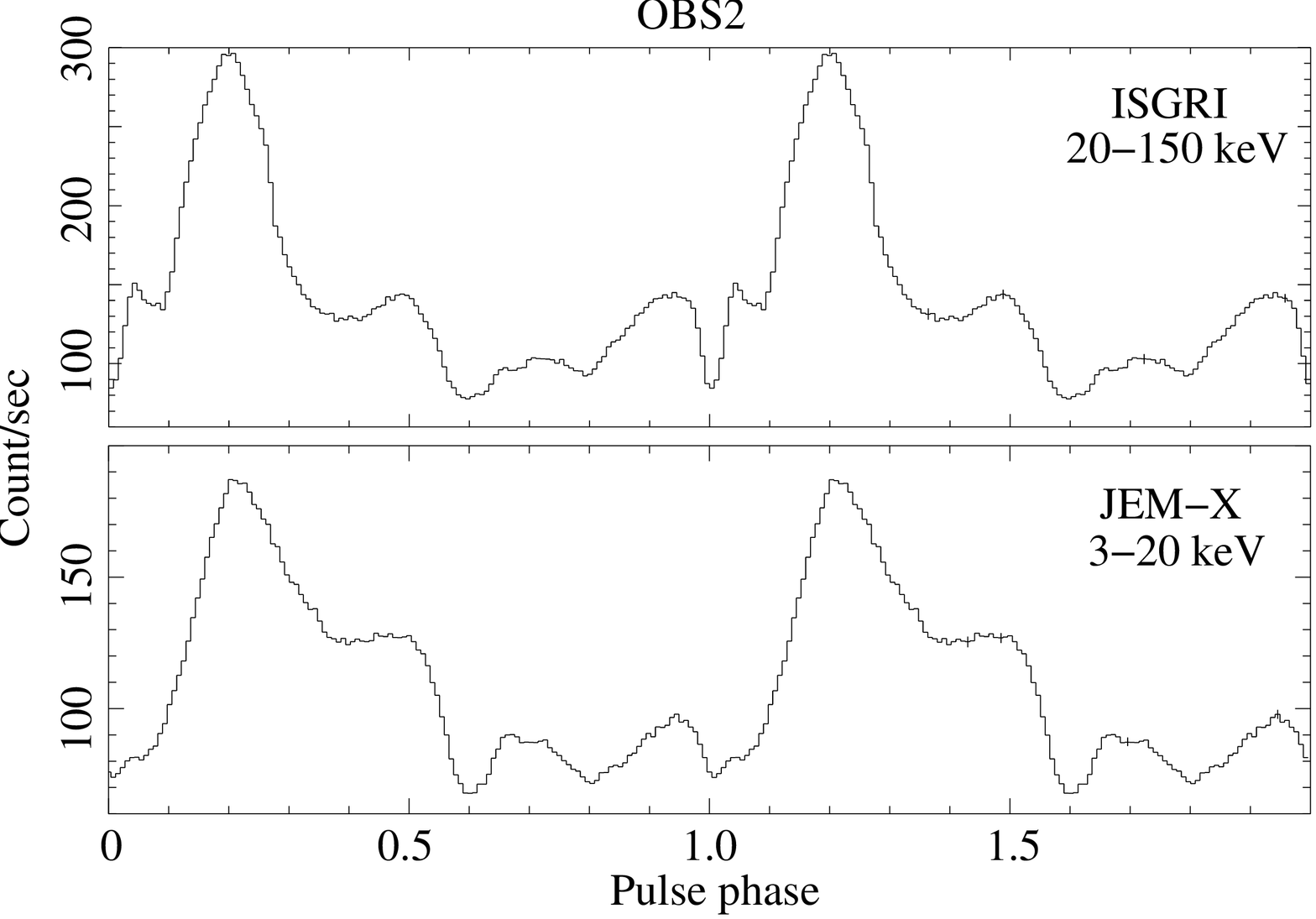}
\caption{X-ray pulse profiles obtained with \textsl{ISGRI} ({\em top panel})
and \textsl{JEM-X} ({\em bottom panel}) during OBS2
(MJD$\sim 53966$), near the maximum of the outburst. 
Phase 0.0 was arbitrarily chosen to be defined by 
the dip-like feature before the main peak.
Typical error bars are indicated.}
\label{pp}
\end{figure}

\begin{figure}
\centering
\includegraphics[width=5.9cm,angle=-90]{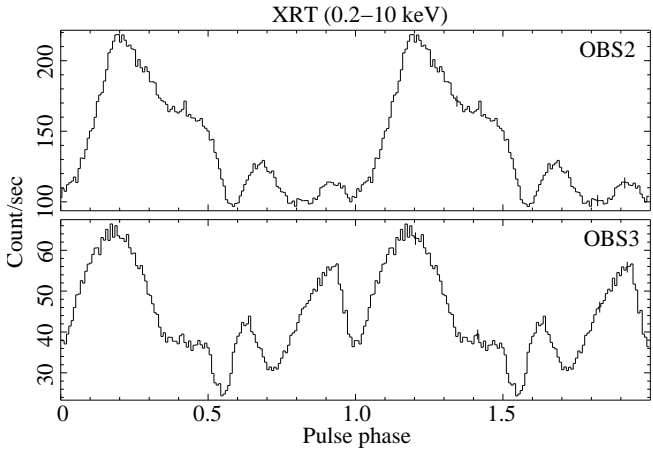}
\caption{X-ray pulse profile obtained with \textsl{XRT} during OBS2
({\em top panel}) and during OBS3 ({\em bottom panel)}.
Typical error bars are indicated.}
\label{xrt}
\end{figure}

Using the values of $P$ and $\dot P$
determined here (Table~\ref{tabper})
we constructed pulse profiles for each observation.
Pulse profiles from OBS1 and OBS2 are found to be very
similar. This can be expected considering that the luminosity of
the source during these observations was approximately the same.
The results are shown in Figs.~\ref{pp} and~\ref{xrt}.
We do not show pulse profiles from OBS1 since there are similar to OBS2.

Fig.~\ref{xrt} shows the 0.2--10 keV pulse profiles
obtained with \textsl{XRT}. One can see that the pulse profile obtained
in OBS3, during the decay of the outburst,  appears
different from that obtained in OBS2 when the luminosity was higher.

The analysis of the pulse profiles, including the comparison
with those obtained during the 1985 outburst and during normal outbursts, is
on-going.

\subsection{Spectral analysis}

For OBS1 pulse averaged X-ray spectra of \textsl{ISGRI} 
and \textsl{JEM-X} were analyzed.
For OBS2 spectra from \textsl{ISGRI, JEM-X, XRT} and \textsl{BAT}
were fitted simultaneously. In order to account for small scale 
uncertainties in the response matrices of the respective instruments
systematic uncertainties were added at a level of
1\% for \textsl{ISGRI} and 3\% for \textsl{JEM-X} and \textsl{XRT}. 
For \textsl{BAT} we used energy-dependent systematic errors provided 
by the \textsl{HEADAS} package.
We note that there are additional global systematic uncertainties 
with the instruments on \textsl{INTEGRAL} (e.g. the canonical spectrum of
the Crab is not reproduced). The stated uncertainties (e. g. in Tables 3
and 4) are therefore to be taken as lower limits.

\begin{table}[b]
\caption{Best fit spectral parameters of EXO 2030+375 using the model
with a ``bump'' around 15 keV (see text).
1$\sigma$(68\%)-uncertainties for one parameter of interest are shown.}
\label{par_bump}
\centering
\renewcommand{\arraystretch}{1.2}
\begin{tabular}{l c c}
\hline\hline
             & 53943(OBS1)    & 53967(OBS2)   \\
Time of observation  & ISGRI, & ISGRI, JEM-X  \\
(MJD) / Instruments  & JEM-X  &   XRT,BAT \\
\hline
$\Gamma$                 & $1.93 \pm 0.01$      & $1.93_{-0.02}^{+0.01}$ \\
$E_{\rm cutoff}$ [keV]   & $25.9_{-0.3}^{+0.2}$ & $26.4_{-0.5}^{+0.2}$\\
$E_{\rm fold}$ [keV]     & $26.1\pm 0.2$        & $26.9\pm 0.2$  \\
$E_{\rm Fe}$ [keV]       & ---                  & $6.6 \pm 0.1$ \\
$\sigma_{\rm Fe}$ [keV]  & ---                  & $1.2\pm 0.1$ \\
$E_{\rm bump}$ [keV]     & $15.3\pm 0.2$        & $13.4_{-0.5}^{+0.4}$\\
$\sigma_{\rm bump}$ [keV]& $2.7 \pm 0.2$        & $4.1_{-0.5}^{+0.3}$   \\
$N_\text{H}$ [$10^{22}{\rm cm}^{-2}$]& $3.1\pm 0.2$ & $3.4\pm 0.1$   \\
$F_\text{JEM-X}$         &      $1.0$\,(fixed)  & $1.0$\,(fixed)   \\
$F_\text{ISGRI}$         &      $1.20\pm 0.01$  & $1.20\pm 0.01$   \\
$F_\text{XRT}$           &       ---            & $0.926\pm 0.004$ \\
$F_\text{BAT}$           &       ---            & $0.91 \pm 0.01$  \\
$\chi^2_\text{red}$/d.o.f.& 1.1/259             &   1.3/937    \\
\hline
\end{tabular}
\end{table}

\begin{figure}
\centering
\includegraphics[width=9.2cm,angle=-90]{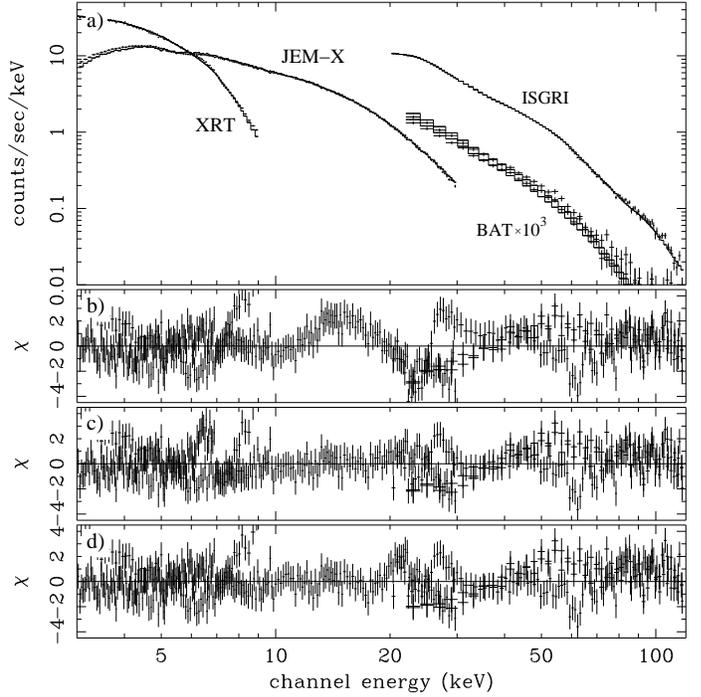}
\caption{The broad band spectrum of EXO 2030+375 from simultaneous fits 
of \textsl{INTEGRAL} and \textsl{Swift} data of OBS2 with \texttt{highecut} (a)
and residual plots after fitting it without additional features (b),
adding a ``bump'' around 15 keV (c), or alternatively 
including two absorption lines at $\sim$10 and $\sim$20 keV (d).}
\label{spe}
\end{figure}

To fit the broad band continuum of the source we used an absorbed
power law with an exponential cutoff (XSPEC \texttt{highecut} model):
\begin{equation*}
\frac{dN}{dE} \propto
\begin{cases}
E^{-\Gamma}, & \text{if\,} E \leq E_{\rm cutoff} \\
E^{-\Gamma} \cdot \exp{\frac{E-E_{\rm cutoff}}{E_{\rm fold}}}, & \text{if\,}
E > E_{\rm cutoff},
\end{cases}
\end{equation*}
where $dN/dE$ is the differential count rate, 
$E$ is the energy of the photons; $\Gamma,\,E_{\rm cutoff}$, 
and $E_{\rm fold}$ are model parameters.
An iron emission line at $\sim$6.5 keV was observed during
OBS2 (which has the largest total number of counts) and modeled by a Gaussian
emission line. \textsl{JEM-X} data revealed an additional feature 
in 10--20 keV range, 
which could be explained by a Gaussian emission model 
at $\sim$13--15 keV 
(a ``bump''). Alternatively, following the suggestion of a
cyclotron line at $\sim$10~keV by \citet{WilsonFinger06}, 
we equally well fitted the spectra adding two Gaussian absorption
lines at $\sim$10 and $\sim$20 keV to the broad band continuum. 
To account for large systematic uncertainties in the absolute flux
measured by the instruments we introduced in our models a free multiplicative factor for each instrument: 
$F_{\rm ISGRI},\,F_{\rm XRT}$, and $F_{\rm BAT}$ (for \textsl{JEM-X} the
factor was fixed to 1.0). The best
fit spectral parameters for OBS1 and OBS2 are listed in
Table~\ref{par_bump} (for the model with a ``bump'') and
Table~\ref{par_lines} (for the model with two absorption lines). 
The broad band spectrum of OBS2 
fitted with \texttt{highecut} is shown in Fig.~\ref{spe}a. 
Figures~\ref{spe}b, ~\ref{spe}c and ~\ref{spe}d show the residuals
after fitting the spectrum by the \texttt{highecut} model without
additional features between 10 and 20 keV, with a ``bump'' around 15\,keV, 
and, alternatively, with two Gaussian absorption lines.

\begin{table}[b]
\caption{Best fit spectral parameters of EXO 2030+375 using the model
with two absorption lines at $\sim$10 and $\sim$20 keV (see text).
1$\sigma$(68\%)-uncertainties for one parameter of interest are shown.}
\label{par_lines}
\centering
\renewcommand{\arraystretch}{1.2}
\begin{tabular}{l c c}
\hline\hline
             & 53943(OBS1)    & 53967(OBS2)   \\
Time of observation  & ISGRI, & ISGRI, JEM-X  \\
(MJD) / Instruments  & JEM-X  &   XRT,BAT \\
\hline
$\Gamma$                     & $1.79\pm 0.01$       & $1.78\pm 0.01$ \\
$E_{\rm cutoff}$ [keV]       & $19.8\pm 0.4$        & $19.5\pm 0.3$\\
$E_{\rm fold}$ [keV]         & $24.2\pm 0.2$        & $24.7_{-0.2}^{+0.1}$  \\
$E_{\rm Fe}$ [keV]           &     ---              & $6.51 \pm 0.03$ \\
$\sigma_{\rm Fe}$ [keV]      &     ---              & $0.32_{-0.03}^{+0.04}$ \\
$E_{\rm 1st~line}$ [keV]     & $10.0_{-0.3}^{+0.2}$ & $10.6_{-0.2}^{+0.1}$\\
$\sigma_{\rm 1st~line}$ [keV]& $1.6_{-0.3}^{+0.4}$  & $0.7 \pm 0.2$   \\
$E_{\rm 2nd~line}$ [keV]     & $20.7\pm 0.3$        & $20.6 \pm 0.3 $\\
$\sigma_{\rm 2nd~line}$ [keV]& $1.3\pm 0.3$  & $2.0_{-0.2}^{+0.3}$   \\
$N_\text{H}$ [$10^{22}{\rm cm}^{-2}$]& $2.2\pm 0.2$ & $3.1\pm 0.1$ \\
$\chi^2_\text{red}$/d.o.f.& 1.1/256                 &   1.3/934    \\
$F_\text{JEM-X}$         &      $1.0$\,(fixed)  & $1.0$\,(fixed)   \\
$F_\text{ISGRI}$             &   $1.25\pm 0.01$     & $1.20\pm 0.01$\\
$F_\text{XRT}$               &          ---         & $0.926\pm 0.004$ \\
$F_\text{BAT}$               &          ---         & $0.90 \pm 0.01$\\
\hline
\end{tabular}
\end{table}

\section{Summary and Discussion}

In June--September 2006 EXO 2030+375 has undergone the second
giant (type II) outburst since its discovery in 1985. During the
outburst the source has been observed with \textsl{INTEGRAL} and
\textsl{Swift}. The main results of these observations can be
summarized as follows:

1. The spin frequency of the pulsar has dramatically increased,
indicating the presence of an accretion disk around the neutron
star. The maximum value of $\dot P$ is found to be comparable to
that observed during the first outburst during which the source was
discovered. The values of $\dot P$ determined for the three
observation periods are linearly related to the X-ray luminosity, 
in line with previous observations \citep{Parmar_etal89a} and the 
prediction of accretion torque theory \citep{GhoshLamb79}.

Since the luminosity of the source and its spin-up rate 
could be measured, we can estimate the value of the 
magnetic field strength on the neutron
star using the classical model of an accreting X-ray pulsar
\citep{PringleRees72}. The equation determining the change of the
angular momentum of the neutron star has the form
$I\dot \omega = \dot M\sqrt{GMR_{\rm A}}$,
where $I$ is the moment of inertia of the neutron star, $\omega$ is
the angular frequency, $R_{\rm A}$ is the Alfv\'en radius, $M$ is
the mass of the neutron star, and $\dot M$ is the accretion rate.
$R_{\rm A}$ depends on the magnetic field strength and the accretion
rate as
$R_{A} \simeq [\mu^2/(2\dot M\sqrt{2GM})]^{2/7}$,
where $\mu$ is the magnetic dipole moment of the neutron star. Using
the observed values of the luminosity $L$ (which can be translated
into the accretion rate using the relation $L \sim 0.1\dot M c^2$),
$\dot P$ and standard neutron star parameters, we obtain an estimate
of the magnetic field strength $B \sim 1-4\cdot 10^{12}$\,Gauss. 
This would be consistent with a fundamental cyclotron 
line in the range $E_{\rm cyc}\sim 10-50$~keV 
($E_{cyc}=11.6\cdot (B/10^{12}\,\text{G})\times[1+z]^{-1}$~keV, 
where $z$ is the gravitational redshift which was assumed to be 0.2 in our 
calculations).

2. For the first time the broad band X-ray spectrum of the source 
(including energies above $\sim$25 keV) during a type II outburst
is analyzed.
The continuum is modeled by a power law with an
exponential cutoff. An iron emission line is observed at $\sim$6--7\,
keV. The spectrum shows additional features between 10 and 20 keV which can
be equally well modeled with a broad emission line at $\sim 13-15$\,keV 
or two absorption lines at $\sim$10 and $\sim$20 keV. On the basis
of phase averaged analysis we can not distinguish
between the two options and we can not confirm the presence
of a cyclotron line at $\sim$36 keV reported by \citet{ReigCoe99}. A
preliminary pulse phase resolved analysis shows that the spectrum is
highly variable with pulse phase. This means that any
interpretation of the phase averaged spectrum should be made
with caution. Moreover, residual features around 25~keV may hint at
not yet well understood calibration uncertainties.

3. The shape of the profiles is found to vary during the outburst, 
probably related to the luminosity.

Pulse-phase-resolved analysis along with the analysis of the
spectral and timing behaviour of the source at different luminosity
states, is on-going.

\begin{acknowledgements}
D.K. thanks Konstantin Postnov for useful suggestions. 
We thank the referee Dr. Lucien Kuiper for his comments and corrections 
which improved clearness and readibility of the paper.
We wish to thank the \textsl{Swift} PI, Neil Gehrels, 
and the Staff of the Swift
Missions Operation Center as well as the \textsl{INTEGRAL} Mission Scientist,
C. Winkler, and the ESA ISOC personnel for their patient help in
scheduling the simultaneous observations of this Target of
Opportunity Program. This research is supported by the German Space
Agency (DLR) under contracts nos. 50 OG 9601 and 50 OG 0501.

\end{acknowledgements}

\bibliographystyle{aa}
\bibliography{6801}

\end{document}